\documentclass[twocolumn,showpacs,preprintnumbers,nofootinbib,nobibnotes,amsmath,amssymb,aps,prl]{revtex4-1}

\usepackage{todonotes}
\usepackage{color}

\usepackage{graphicx}
\usepackage{hyperref}
\usepackage{slashed}

\newcommand{\cL}{\mathcal{L}}
\newcommand{\cH}{\mathcal{H}}

\newcommand{\res}{\mathrm{res}}
\newcommand{\wres}{\omega_\mathrm{res}}
\newcommand{\eV}{\text{\,eV}}
\newcommand{\keV}{\text{\,keV}}
\newcommand{\MeV}{\text{\,MeV}}
\newcommand{\GeV}{\text{\,GeV}}
\newcommand{\disp}{\displaystyle}

\renewcommand{\l}{\left(}
\renewcommand{\r}{\right)}

\newcommand{\lAngle}{\langle\!\langle}
\newcommand{\rAngle}{\rangle\!\rangle}

\begin{document}

\preprint{INR-TH-2018-022}

\title{Induced resonance makes light sterile neutrino Dark Matter cool}

\author{F. Bezrukov}
\email{Fedor.Bezrukov@manchester.ac.uk}
\affiliation{The University of Manchester, School of Physics and Astronomy,\\
  Oxford Road, Manchester M13 9PL, United Kingdom}
\author{A. Chudaykin}
\email{chudy@ms2.inr.ac.ru}
\affiliation{Institute for Nuclear Research of the Russian Academy of Sciences,\\
  60th October Anniversary prospect 7a, Moscow 117312, Russia}
\affiliation{Moscow Institute of Physics and Technology,\\
  Institutsky per. 9, Dolgoprudny 141700, Russia}
\author{D. Gorbunov}
\email{gorby@ms2.inr.ac.ru}
\affiliation{Institute for Nuclear Research of the Russian Academy of Sciences,\\
  60th October Anniversary prospect 7a, Moscow 117312, Russia}
\affiliation{Moscow Institute of Physics and Technology,\\
  Institutsky per. 9, Dolgoprudny 141700, Russia}

\date{\today}

\begin{abstract}
  We describe two new generation mechanisms for Dark Matter composed
  of sterile neutrinos with ${\cal O}(1)$\,keV mass.  The model contains a
  light scalar field which coherently oscillates in the early Universe
  and modulates the Majorana mass of the sterile neutrino. In a region
  of model parameter space, the oscillations between active and
  sterile neutrinos are resonantly enhanced.  This mechanism allows us
  to produce sterile neutrino DM with small mixing angle with active
  neutrinos, thus evading the X-ray constraints.  At the same time the
  spectrum of produced DM is much cooler, than in the case of ordinary
  oscillations in plasma, opening a window of lower mass DM, which is
  otherwise forbidden by structure formation considerations. In other
  regions of the model parameter space, where the resonance does not
  appear, another mechanism can operate: large field suppresses the
  active-sterile oscillations, but instead sterile neutrinos are
  produced by the oscillating scalar field when the effective fermion
  mass crosses zero.  In this case DM component is cold, and even
  1\,keV neutrino is consistent with the cosmic structure formation.
\end{abstract}

\maketitle

\section{Introduction}

One of the major problems of the present day particle physics is to
provide a viable Dark Matter (DM) candidate, which is an important
component of the standard cosmological model.  One of the
well-motivated candidates is sterile neutrino.  A minimal extension of
the Standard Model of particle physics (SM), providing such a
candidate, is $\nu$MSM \cite{Asaka:2005an,Asaka:2005pn}, which extends
SM with three right handed neutrino partners.  If one of the sterile
neutrinos $N$ is light, with mass in ${\cal O}(1-100)$~keV range, it is
stable enough on cosmological time scales to form
DM~\cite{Dolgov:2000ew}.  The simplest model is parametrized by
sterile neutrino mass $M$ and mixing angle with the active neutrino
$\theta_0$ \cite{Adhikari:2016bei}.  Main constraints on the model
arrive from two-body decay $N\to\nu\gamma$, which leads to X-ray
signal from DM dominated regions of the sky, placing the upper bound
on $\theta_0$.  At the same time, production of the correct DM amount
$\Omega_N=\Omega_{DM}$ of sterile neutrino in the early Universe
requires some sizeable mixing $\theta_0$.  Third constraint arrives
from the investigation of the cosmic structure formation, which limits
the \emph{velocity} distribution of DM particles.  For light DM
particle this results in a significant constraint from below on the
mass, which depends on the production spectrum (note, that such a
light DM can never equilibrate with primordial plasma).  Thus
production mechanisms with cold or mildly warm spectrum are required.

Accommodating all these constraints is a non-trivial task.  The
simplest mechanism of production in active-sterile oscillations
\cite{Dodelson:1993je} contradicts either X-ray or structure formation
constraints.  Production in lepton-asymmetric plasma is resonantly
enhanced~\cite{Shi:1998km} and hence works at smaller mixing angle
providing with colder spectrum. Still it requires at least
$M\gtrsim5-7\keV$~\cite{Baur:2017stq} and fine-tuning of the model
parameters.  A number of production mechanisms were proposed to generate the colder neutrino evading the structure formation constraints, see \cite{Adhikari:2016bei} for a
review. 


In this letter two new mechanisms of production of the DM sterile
neutrino are proposed, both operating in the presence of a coherently
oscillating Majoron type scalar field in the early Universe.  Sterile
neutrino dynamics of such a system is rather involved.  When the
effective Majorana mass induced by the scalar background is large, no
sterile neutrinos are generated from the thermal bath of active
neutrinos.  At the same time, periodic crossing of zero Majorana mass
$M(t)=0$ leads to non-trivial resonance effects \emph{both} in the
active-sterile oscillations (described in the next section) \emph{and}
direct production by the scalar field (effective for different
parameter choice and outlined in the third section).
The first mechanism gives the coldest sterile neutrino distribution compared to all other mechanism relying on active-sterile oscillations and works for small active-sterile mixing angles.  The direct production mechanism provides the coldest spectrum for sterile neutrino produced without relation to active-sterile neutrino mixing.


\section{Induced neutrino resonance}

The two main components of the model are the right handed sterile
neutrino $N$ and the light scalar field $\phi$,
\begin{equation}
  \label{lagr}
  \begin{split}
  \cL =
  &\ i\bar{\nu}\slashed{\partial}\nu + i\bar{N}\slashed{\partial}N
    + \frac{1}{2}(\partial_\mu\phi)^2 + \frac{m_\phi^2}{2}\phi^2 \\
  & + (m_D\bar{\nu}N
    + \frac{M_0}{2}\bar{N}^cN + \frac{f}{2}\phi\bar{N}^cN+\mathrm{h.c.}),
\end{split}
\end{equation}
where $\nu$ is an active SM neutrino, and the Majorana mass is much
larger than the Dirac mass, $M_0\gg m_D$.  At present the scalar field
is in the vacuum $\phi=0$, and diagonalization of the mass term in
\eqref{lagr} yields mixing between heavy (sterile) and light (active)
neutrino,
\begin{equation}
  \label{eq:theta0}
  \theta_0 \simeq m_D/M_0.
\end{equation}
We assume, that the scalar is oscillating coherently in the early
Universe with amplitude decaying with plasma temperature as
$\propto T^{3/2}$. This induces the temperature(time)-dependent
Majorana mass for $N$
\begin{equation}
  \label{mass}
  M(t) = M_0+M_A\sin m_\phi t, \quad
  M_A\equiv M_0\left(
    \frac{g_* T^3}{g_{*,e}^{\vphantom{3}} T_e^3}
  \right)^{\!1/2}\!\!\!,
\end{equation}
where $T_e$ is the temperature when the amplitude of oscillating part
in \eqref{mass} equals the bare Majorana mass $M_A=M_0$.  We ignore
any difference between the effective numbers of degrees of freedom in
plasma, $g_*=g_*(T)$, which enter energy and entropy densities. Note,
that the time scale of scalar oscillations $m_\phi^{-1}$ is much
shorter than that of the change of the amplitude $M_A$, which is
$2/(3H)$ with the Hubble parameter $H$ characterizing the Universe
expansion rate.

\begin{figure}
  \includegraphics[width=\linewidth]{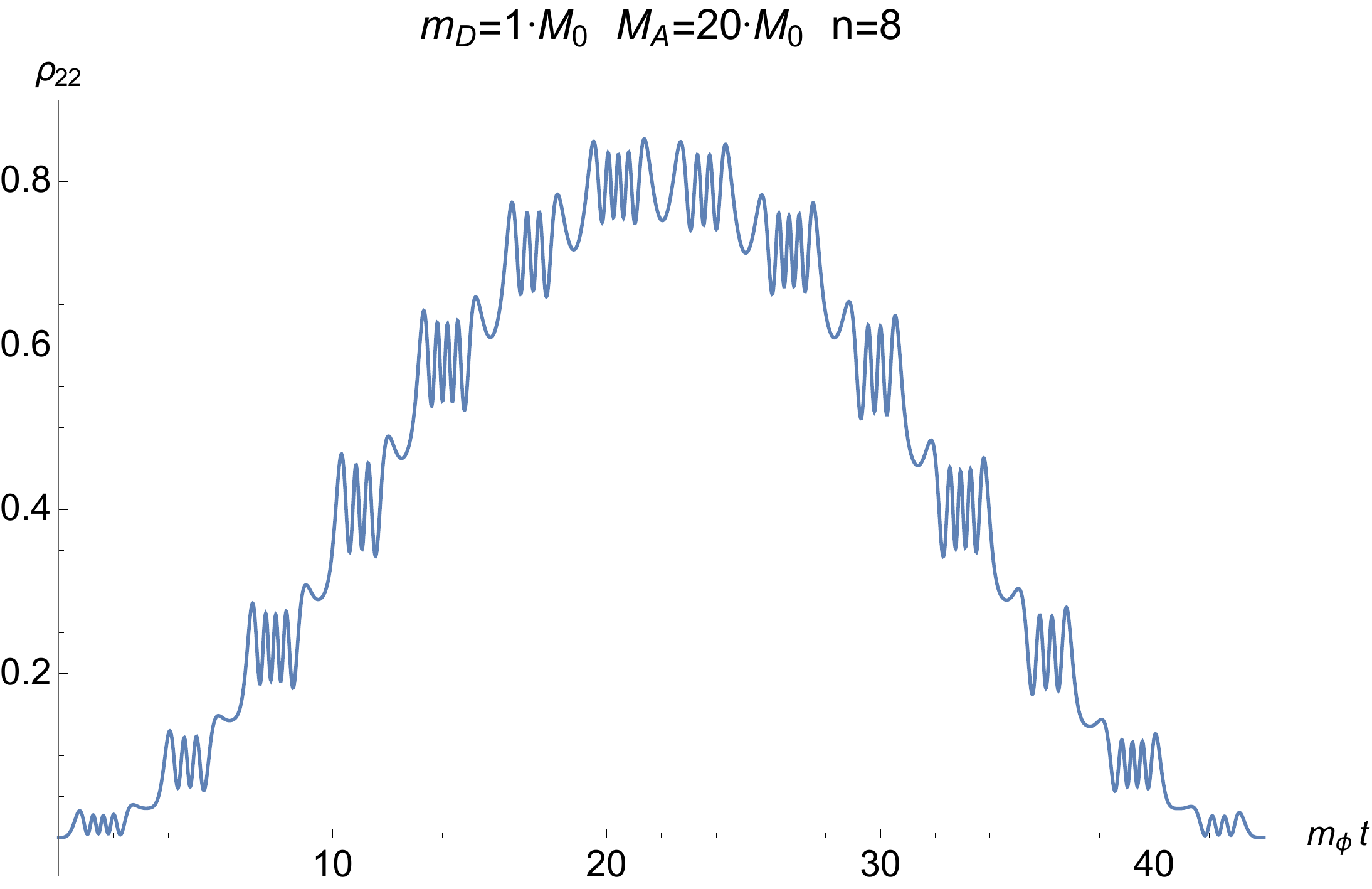}
  \caption{Example of resonance evolution of transition probability at
    $n=8$. Relation between $M_0$ and $p$ is defined
    by\,\eqref{eq:res}. Fast oscillations with rate $M_A^2/4p\approx8m$,
    slower oscillations with rate $m$ related to the scalar field
    evolution, and resonant modulation by frequency~\eqref{eq:wres}
    are shown here.}
  \label{fig:res}
\end{figure}

The coherent evolution of the neutrino states of given momentum $p$
(we also use dimensionless conformal momentum $y\equiv p/T$ when
convenient) is described (for relativistic momenta
$p\simeq 3T\gg M(t)$) by the equation
\begin{equation}\label{eq:osceq}
  i\frac{\partial}{\partial t} \rho =
  [\cH, \rho]-\frac{i}{2} \{ \Gamma, \rho-\rho_\mathrm{eq}\}.
\end{equation}
on the 2$\times$2 density matrix $\rho$ corresponding to active and
sterile neutrinos, with $\rho_{11}$ ($\rho_{22}$) being the
probability density of the active (sterile) neutrino.  Here
$\rho_\mathrm{eq}=\mathrm{diag}(f_\mathrm{FD}(y), f_\mathrm{FD}(y))$
is the equilibrium Fermi-Dirac distribution and
$\Gamma=\left( \begin{array}{cc} \Gamma_A & 0 \\ 0 &
    0\end{array}\right)$ is the damping term due to active neutrino
interactions in the thermal bath.
The Hamiltonian of the system in the plasma reads
\[
\cH=\frac{\Delta_0}{2}\left(\begin{array}{cc}
                              -\cos2\theta & \sin2\theta\\
                              \sin2\theta & \cos2\theta
          \end{array}
        \right),\;\text{where}\quad \Delta_0 = \frac{\Delta m^2}{2 p}
\]
with time-dependent oscillation parameters 
\[
  \Delta m^2 = M\sqrt{M^2+4m_D^2}\,,\;\;
  \tan\theta  = \frac{2 m_D}{M+\sqrt{M^2+4 m_D^2}}.
\]
Equation \eqref{eq:osceq} exhibits a non-trivial resonance behaviour
for neutrino momentum $p_\res$ if the typical (properly averaged)
frequency of neutrino oscillations is an integer multiple of the
frequency of the scalar background oscillations $m_\phi$,
\begin{equation}
  \label{eq:res}
  \frac{M_A^2+2M_0^2}{4p_\res} = n m_\phi,
  \quad n\in \mathbb{N}.
\end{equation}
It can be shown \cite{LongScalarRes} that in the resonance in the
absence of scattering, $\Gamma_A=0$, (and with odd $n$) the $\nu-N$
oscillations proceed effectively (see Fig.~\ref{fig:res}) with maximum
mixing $\rho_{22}\simeq \sin^2(\wres t/2)$ at frequency
\begin{equation}
  \label{eq:wres}
  \wres \simeq
  1.3 m_\phi\frac{m_D}{M_A}n^{1/3}\cos\frac{4M_0}{M_A}n
  \simeq 1.3 m_\phi\frac{m_D}{M_A} n^{1/3},
\end{equation}
which is a good approximation for $M_A\gg M_0$.  Numerically it works
well until $M_A\gtrsim 5M_0$.

Note that for our choice of parameters this frequency is much lower
than both other typical oscillation frequencies $m_\phi$ and
$M_A^2/4p$.  The resonance is narrow in momentum, with typical width
\begin{equation}
  \label{eq:dpres}
  \frac{\Delta p_\res}{p_\res} \simeq
  \frac{\sqrt{2}\wres}{n m_\phi}\simeq\frac{2}{n}\frac{m_D}{M_A}\ll 1.
\end{equation}
We focus at the resonance with the highest momentum, that is $n=1$.
Higher resonances are relevant at lower neutrino momentum, while for
the purpose of this work the higher momentum region is interesting,
because it gives the most important contribution both to the neutrino
abundance and to the average neutrino velocity.

For the resonant frequencies the amount of sterile neutrino reaches
Fermi-Dirac distribution after typical time $\wres^{-1}$.  However,
there are two effects that can prevent the resonance \eqref{eq:wres}
from happening.  First, scattering of active neutrino in plasma with
$\Gamma_A>\wres$ slows down the conversion and increases the resonance
width \cite{LongScalarRes}.  Second, due to the Universe expansion the
conformal momentum $y$ of the resonance changes with temperature and
$M_A$ \eqref{eq:res}.  At lower temperatures lower $y$ enter into
resonance.  However, if the whole resonance band of the width
$\Delta y_\res=\Delta p_\res/T$ \eqref{eq:dpres} moves through a given
frequency $y$ faster, than the duration of one oscillation
$\wres^{-1}$, the resonance becomes ineffective (``narrow'').  This
happens at high temperatures when
\begin{equation}
  \label{eq:condH}
  \frac{\Delta y_\res}{y_\res}\frac{y_\res}{\dot y_\res} \wres
  = \frac{\wres^2}{\sqrt{2}m_\phi H} \lesssim 1,
\end{equation}
where $H=T^2/M_{\mathrm{Pl},*}$. Equation \eqref{eq:condH} defines the
temperature above which the resonance is too narrow, and the sterile
neutrino distribution function is suppressed, as compared to the
Fermi-Dirac distribution.

\subsection{Spectrum with narrow resonance}

\begin{figure}[!htb]
  \includegraphics[width=\linewidth]{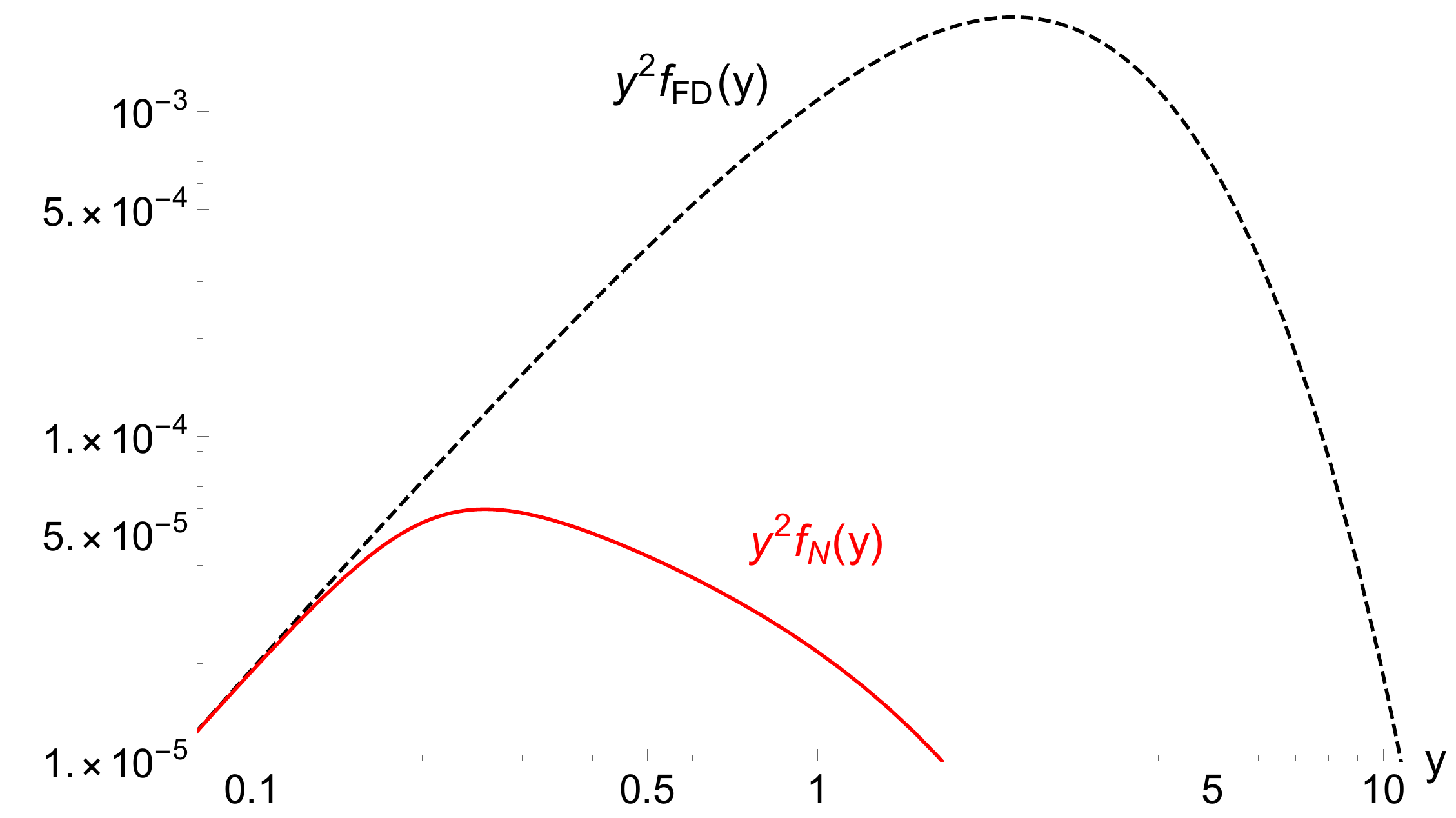}
  \caption{Fermi-Dirac $f_\mathrm{FD}$ and sterile neutrino $f_N$ spectra in logarithmic scale.}
  \label{fig:spect}
\end{figure}

The sterile neutrino spectrum $f_N$ is given by the Fermi-Dirac
distribution for low momenta, and is cut off at $y_s$ saturating the
condition \eqref{eq:condH}.  One finds approximately
\cite{LongScalarRes}
\begin{equation}\label{spect_y2}
  f_N(y) = \frac{f_\mathrm{FD}(y)}{\sqrt{1+0.8\left(\frac{\disp y}{\disp
          y_s}\right)^5}}.
\end{equation}
The spectrum is suppressed at high momenta, see Fig.~\ref{fig:spect}.

Cut-off value $y_s$ defines the neutrino abundance today
$\Omega_N\rho_\mathrm{crit}=M_0 2\frac{4}{11}T_0^3\int4\pi
y^2f_N(y)dy$. The DM abundance $\Omega_N=\Omega_{DM}$ is achieved in
the mass range $1\keV\lesssim M_0\lesssim 20\keV$ for
\begin{equation}
  \label{eq:yg2}
  y_s\simeq 0.2\left(\frac{1\keV}{M_0}\right)^{2/5}.
\end{equation}
This mode is resonant at the temperature
\begin{align}\begin{split}
  \label{eq:Tg2}
  T_s = 3T_e
  \left(\frac{T_e}{11\MeV}\right)^{1/2}
  \left(\frac{m_\phi}{1\eV}\right)^{1/2}\\
  \left(\frac{1\keV}{M_0}\right)^{6/5}
  \left(\frac{g_{*,e}}{g_{*,s}}\right)^{1/2}.
\end{split}\end{align}
The correct DM neutrino abundance is achieved if the resonance becomes
too narrow \eqref{eq:condH} at this temperature, leading to the final
result for the mixing angle \eqref{eq:condH}
\begin{align}\begin{split}
  \label{eq:thetani2}
  \theta_0 \sim 3.2\times10^{-6}
  \left(\frac{1\keV}{M_0}\right)^{3}
  \left(\frac{g_{*,s}}{10.75}\right)^{1/4}
  \left(\frac{m_\phi}{1\eV}\right)^{3/4}\\
  \times
  \left(\frac{T_e}{11\MeV}\right)^{9/4}.
\end{split}\end{align}
The dark matter sterile neutrino spectrum \eqref{spect_y2} is cool
with average 
\begin{equation}\label{y}
\lAngle y\rAngle=1/\mathcal{S}^{1/3}
\end{equation} 
where dilution factor $\mathcal{S}=g_{*,s}/10.75$ accounts for the
amount of entropy that was released after sterile neutrino production.

For conservative choice $\mathcal{S}=1$ (corresponding to the
parameter choice in this section) the Ly-$\alpha$ constraint
$m_\mathrm{NRP}>8\keV$~\cite{Adhikari:2016bei} is translated in our
case to
\begin{equation}\label{massFS}
M_0>2.5\keV
\end{equation} 

The main advantage of the proposed generation mechanism is a cold
sterile neutrino spectrum \eqref{spect_y2} which opens a new window
for warm dark matter. In what follows we compare in different
  scenarios average momenta of DM particles calculated just before
  active neutrino freezout, corresponding to $g_*=10.75$.  Two common
mechanisms that rely on the active-sterile neutrino mixing provide a
warmer spectrum: near thermal distribution for non-resonant production
with $\lAngle y\rAngle_{DW}=3.15$ \cite{Dodelson:1993je}, and resonant
production in presence of lepton asymmetries can lead to average
momenta as low as $\lAngle y\rAngle_{SF}=1.8/\mathcal{S}^{1/3}$
\cite{Abazajian:2001nj}.  There is a wider class of mechanisms that do
not rely on active-sterile neutrino mixing for the DM generation.  The
notable examples are decays of thermalised particles leading to
$\lAngle y\rAngle_{D}=2.45/\mathcal{S}^{1/3}$ where dilution factor
$\mathcal{S}$ depends on production time of sterile neutrinos and is
unique for each individual scenario
\cite{Shaposhnikov:2006xi,Kusenko:2006rh} \footnote{Frequently, 
the  average momenta at $T\ll1\MeV$ are instead given. In this case,
  decays of thermalised particles at $T\simeq 100\GeV$ leads to
  $\lAngle y\rAngle_{D,T\ll1\MeV}=0.81$ whereas our result \eqref{y}
  is rescaled as $\lAngle y\rAngle_{T\ll1\MeV}=0.71$. We note that
  value $g_*=3.36$ used in \cite{Kusenko:2006rh,Petraki:2007gq} should
  be replaced with $g_*=43/11\approx3.9$, see
  \cite{Bezrukov:2014qda}.}.  There are more possibilities if the
scalar is extremely weakly coupled and freezes-out before decaying to
sterile neutrinos, generally leading to a slightly warmer DM
\cite{Petraki:2007gq,Petraki:2008ef}, unless the decay happens at high
temperatures \cite{Merle:2013wta}.  Only the models with a significant
entropy release from some beyond SM processes after the sterile
neutrino production \cite{Bezrukov:2009th,Kusenko:2010ik} lead to
colder spectrum with $\lAngle y\rAngle=0.7$.

\subsection{Discussion of the parameter region}

\begin{figure}[!htb]
  \includegraphics[width=\linewidth]{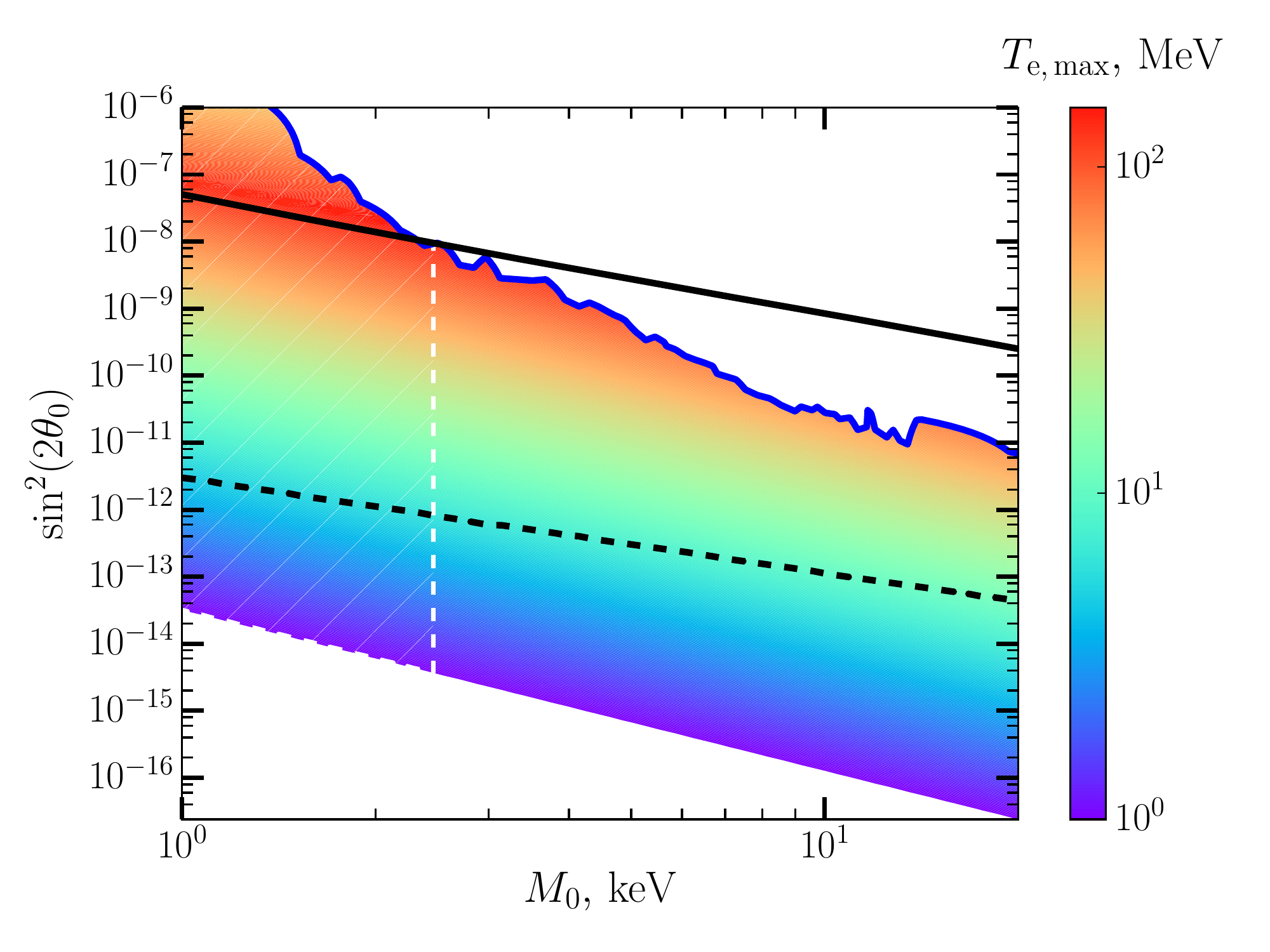}
  \caption{Parameter scan in the sterile neutrino mass- mixing angle
    plane (the corresponding scalar masses
      $10^{-2}\eV<m_\phi<M_0$ help to avoid additional production of
      sterile neutrinos from scalar decays at $T\simeq m_\phi$ as in
      Refs.\    \cite{Shaposhnikov:2006xi,Kusenko:2006rh,Petraki:2007gq}).  Blue
    line is the upper limit from X-ray
    observations~\cite{Adhikari:2016bei}.  Colour region gives the
    fraction of the sterile neutrino equal to DM
    $\Omega_N=\Omega_{DM}$. Colour indicates the maximal value $T_e$
    satisfying $T_e<T_s/3$ and not leading to overproduction of dark
    matter via non-resonant generation at $T\lesssim T_e$.  White
    dashed region is excluded by studying the cosmic structure
    formation. For reference: black line is for the conventional
    non-resonant generation mechanism~\cite{Adhikari:2016bei}, black
    dashed line corresponds to the maximal lepton asymmetry attainable
    in the $\nu$MSM~\cite{Adhikari:2016bei}.}
  \label{fig:region}
\end{figure}

The calculations in the previous subsection are valid if there is a
sufficient region of temperatures for the resonance, that is
$T_s\gtrsim3T_e$ (so that $M_A\gtrsim5M_0$ is true).  For each value
of $(M_0,\theta)$ this requirement limits $T_e$ from above (and
$m_\phi$ from below), see Fig.~\ref{fig:region} and \eqref{eq:Tg2},
\eqref{eq:thetani2}.

The maximum possible mixing angle is defined by the X-ray
observational constraints~\cite{Adhikari:2016bei}.  At low values of
the mixing angle the temperature $T_e$ drops below 1\MeV{}, and with
active neutrinos out of equilibrium the analysis of (\ref{eq:osceq})
is more involved, which defines the lower boundary of the region in
Fig.~\ref{fig:region}.

For large values of $\theta_0$, which are close to the non-resonant
production line, there is also a contribution from simple
active-sterile oscillations at $T<T_e$ (c.f.~\cite{Bezrukov:2017ike})
which can be suppressed by lowering $T_e$.  This leads to additional
constraints at low $M_0$ and high $\theta_0$.

Note, that the production in resonantly-enhanced oscillations depends
on the properties of the scalar sector only via $m_\phi$ and $T_e$,
and not directly on the Yukawa $f$.  The Yukawa depends on the initial
value of the coherently oscillating scalar field.  Thus, it is not
required to have large energy density in the scalar field
condensate--it only serves to induce the resonance transfer between
the active and sterile neutrino sectors.  Sufficiently large (but well
within perturbativity limits, c.f.~\eqref{eq:fforDM}) value of $f$
makes the mechanism described in the next section inactive for
the choice of parameters in Fig.~\ref{fig:region}.  At the same time,
for smaller values of $f$ it may be possible to have the scalar 
contributing to the present DM, leading to multi-component cold+warm
DM with potentially non-trivial structure formation at
small scales.

\section{Production by oscillating scalar}

\begin{figure}[!htb]
  \includegraphics[width=0.95\linewidth]{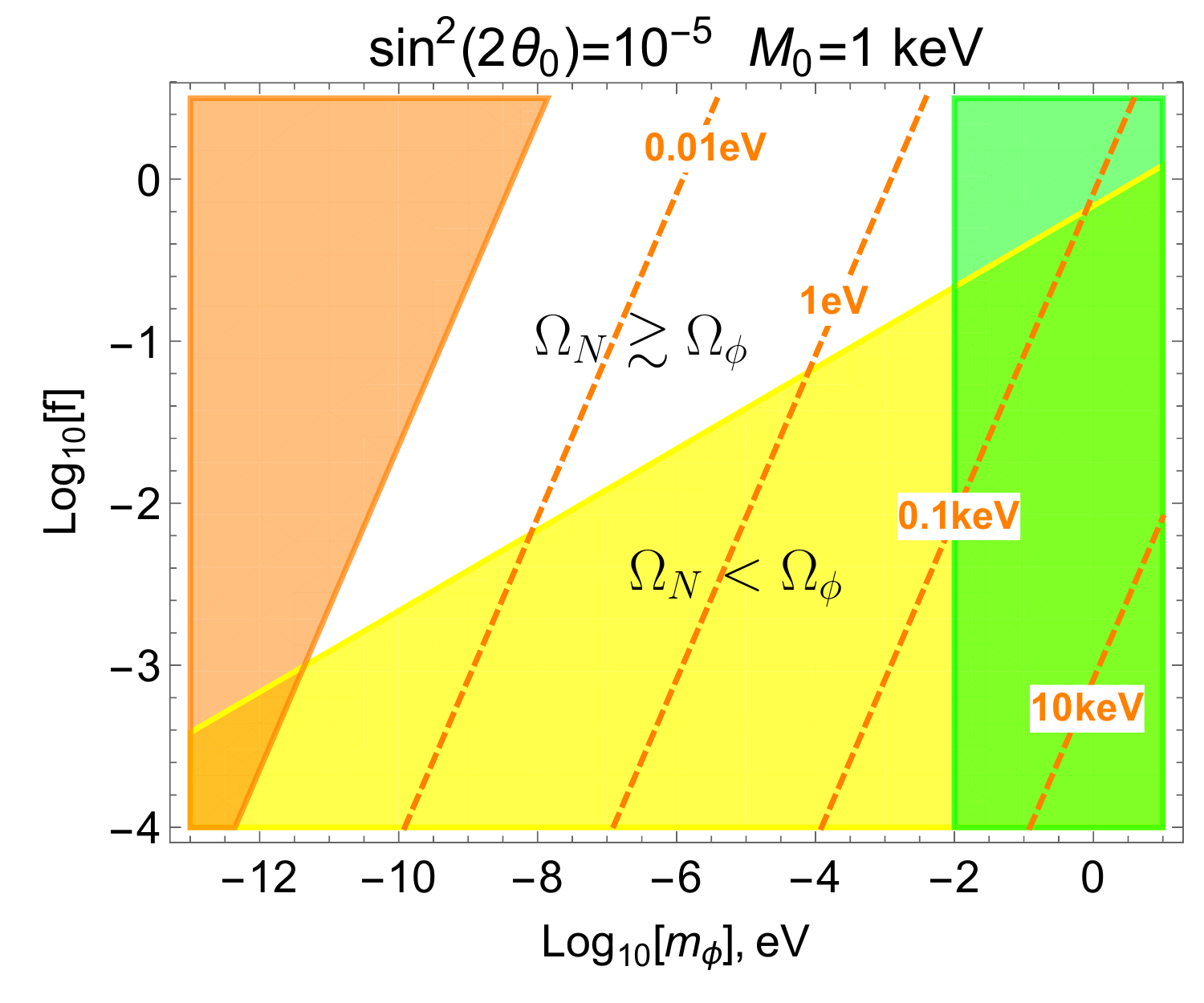}
  \caption{Direct production of the sterile neutrino by the
    oscillating scalar field.  At each point
    $\Omega_{DM}=\Omega_\phi+\Omega_N$ except for the green region
    (right) where~\eqref{scalar-to-active} is violated. Orange region
    (left) is rulled out by~\eqref{numass_today}. White region
    corresponds to the significant contribution of sterile neutrino
    $\Omega_N\gtrsim\Omega_\phi$ whereas yellow field refers to
    $\Omega_N<\Omega_\phi$. Dashed lines correspond to several values
    of $T_e$.}
  \label{fig:Nonpert}
\end{figure}

In the limit of very low $T_e$ (or, alternatively, small $\theta_0$)
neither the mechanism of the last section, nor the usual non-resonant
neutrino oscillations work \cite{Bezrukov:2017ike}. However, for
$M_A>M_0$ ($T>T_e$) \emph{direct} production of the sterile neutrino
by the scalar field is possible at the moments when the mass $M(t)$
crosses zero \cite{Giudice:1999fb,Gorbunov:2011zzc}.  The process
terminates at $T=T_e$ \eqref{mass}, and the density of
\emph{non-relativistic} sterile neutrino produced in this way
reads~\cite{Gorbunov:2011zzc},
$n_N\simeq(2/6\pi^2)(M_0 m_\phi)^{3/2}$.  The field amplitude at that
moment $\phi_e=M_0/f$ \eqref{lagr} controls the scalar energy density,
$\rho_\phi=m_\phi^2\phi_e^2/2$. The scalar field can be stable at
cosmological time-scales because of kinematics, $m_\phi<M_0$, and
sufficiently small mixing yielding for the scalar width
\begin{equation}
  \label{scalar-to-active}
  \frac{\Gamma_{\phi\to\nu\nu}}{H_0} \equiv
  \theta_0^4\times \frac{f^2}{16\pi}\frac{m_\phi}{H_0}\ll 1 .
\end{equation}
Then 
the ratio of energy densities of neutrinos
and scalars remain constant, and their present relative contributions
are related as
\begin{multline}\label{Omega-osc}
  \frac{\Omega_N}{\Omega_\phi}
  =\frac{2\rho_N(T_e)}{m_\phi^2\phi_e^2}\simeq
  0.2\left(\frac{f}{0.1}\right)^2\!\sqrt{\frac{M_0}{1\keV}}\,\sqrt{\frac{0.01\eV}{m_\phi}}
\end{multline}
The total present DM is composed of the mixture of sterile neutrino
and the coherently oscillating scalar
$\Omega_{DM}=\Omega_\phi+\Omega_N$, with the total abundance defined
by the scalar field amplitude at high temperatures, which allows to
find the Yukawa coupling
\begin{equation}
  \label{eq:fforDM}
  f^2=\frac{m_\phi^2M_0^2}{2\Omega_{DM}\rho_\mathrm{crit}}
  \frac{g_{*,0}T_0^3}{g_{*,e}T_e^3}.
\end{equation}
To have $N$ mass constant at present we need $T_e>T_0$, which implies
\begin{equation}\label{numass_today}
  0.4\l\frac{0.1}{f}\r^{\frac{2}{3}}\!
  \l\frac{m_\phi}{10^{-10}\mathrm{eV}}\r^{\frac{2}{3}}
  \l\frac{M_0}{1\mathrm{keV}}\r^{\frac{2}{3}}\!
  \l\frac{g_{*,0}}{g_{*,e}}\r^{\frac{1}{3}}\!>1.
\end{equation} 
This production mechanism works at very low $T_e$, with DM distributed
between the scalar field and sterile neutrino, Fig.~\ref{fig:Nonpert}.

It should be noted, that the exact distribution of density between the
two types of DM is hard to determine for
$\Omega_N/\Omega_\phi\gtrsim1$, as far as the backreaction of the
produced sterile neutrino background on the scalar field becomes
important.  We also stress that neutrino oscillations are efficiently
suppressed owing to the choice of very small $T_e$, see
Fig.~\ref{fig:Nonpert}. Hence not only low, but moderate mixing angles
are allowed (above usual non-resonant generation limit), the upper
limits follow from $X$-ray searches alone.  This option opens new
perspectives for direct searches of sterile neutrinos in the future
\cite{Bezrukov:2017ike}.

We emphasize that the sterile neutrino directly produced by
oscillating background is cold. Indeed, at $T=T_e$, when modes of the
highest momenta $p\lesssim p_f$ are produced,  all the bigger ones are
exponentially suppressed. Since $p_f\sim\sqrt{M_0 m_\phi}\ll M_0$
\cite{Gorbunov:2011zzc}, the generated sterile neutrinos are nonrelativistic. It allows to avoid any structure formation constraints. 

\section{Conclusions}

The presence of a coherently oscillating scalar background affects the
dynamics of sterile neutrinos significantly.  As compared to the
traditional sterile neutrino DM scanarios, this allows either to
suppress the creation of sterile neutrinos from active neutrino
oscillations in plasma, or to enhance these oscillations resonantly.
A mechanism converting directly between the scalar background and
sterile neutrinos is also possible.  In general, combination of all
three mechanisms is active at various stages of the thermal history of
the Universe, allowing for a rich variety of effects, including
generation of Warm DM, generation of sterile neutrino Cold DM
component along with the scalar DM component, and also cosmologically
viable DM sterile neutrinos with relatively large mixing angle.  A
detailed study of these effects (see also \cite{LongScalarRes}),
including more realistic study of the neutrino backreaction on the
scalar and quantum corrections to the scalar dynamics, may reveal
other interesting scenarios.

\vspace{0.2cm}
\begin{acknowledgments}
  The study of resonant production (AC and DG) is supported by RSF
  grant 17-12-01547. The work of FB is supported in part by the
  Lancaster-Manchester-Sheffield Consortium for Fundamental Physics,
  under STFC research grant ST/L000520/1 and by the IPPP associateship.
\end{acknowledgments}

\bibliography{sres-2}

\end{document}